\documentclass[12pt,preprint]{aastex62}
\usepackage{rotating}
\usepackage{epsfig}
\newcommand{\msun}{{ M_\odot }}  
\newcommand{\yr}{{ \rm yr }}

\newcommand{\kms}{{ \rm km~s^{-1} }}   

\begin{document}

\title{Flat Spectrum Radio Continuum Emission Associated with
 $\epsilon$ Eridani}

\author{Luis F. Rodr{\'\i}guez}
\affil{Instituto de Radioastronom\'\i a y Astrof\'\i sica, 
Universidad Nacional Aut\'onoma de M\'exico,, Apdo. Postal 3-72 (Xangari), 58089 Morelia, Michoac\'an, M\'exico}
\affil{Mesoamerican Center for Theoretical Physics, Universidad
Aut\'onoma de Chiapas, Carretera Emiliano Zapata Km.4,
Real del Bosque (Ter\'an). 29050 Tuxtla Guti\'errez, Chiapas,
M\'exico}

\author{Susana Lizano}
\affil{Instituto de Radioastronom\'\i a y Astrof\'\i sica, 
Universidad Nacional Aut\'onoma de M\'exico,, Apdo. Postal 3-72 (Xangari), 58089 Morelia, Michoac\'an, M\'exico}

\author{Laurent Loinard}
\affil{Instituto de Radioastronom\'\i a y Astrof\'\i sica, 
Universidad Nacional Aut\'onoma de M\'exico,, Apdo. Postal 3-72 (Xangari), 58089 Morelia, Michoac\'an, M\'exico}
\affil{Instituto de Astronom\'\i a, Universidad Nacional Aut\'onoma de M\'exico, Apartado Postal 70-264, CdMx, CP 04510, M\'exico}

\author{Miguel Ch\'avez-Dagostino}
\affil{Instituto Nacional de Astrof\'\i sica, Optica y Electr\'onica, Luis Enrique Erro \#1, CP 72840, Tonantzintla, Puebla, M\'exico}

\author{Timothy S. Bastian}
\affil{National Radio Astronomy Observatory, 520 Edgemont Road, Charlottesville, VA 22903, USA}

\author{Anthony J. Beasley}
\affil{National Radio Astronomy Observatory, 520 Edgemont Road, Charlottesville, VA 22903, USA}

\email{l.rodriguez@irya.unam.mx}
 
\begin{abstract}

We present Very Large Array observations at 33.0 GHz that detect emission coincident with $\epsilon$ Eridani to within $0\rlap.{''}07$
(0.2 AU at the distance of this star), with a positional accuracy of $0\rlap.{''}05$. This result strongly supports the suggestion of previous authors that the quiescent centimeter emission
comes from the star and not from a proposed giant exoplanet with a semi-major axis of $\sim1\rlap.{''}0$ (3.4 AU). 
The centimeter emission is
remarkably flat and is consistent with optically thin free-free emission. In particular, it can be modeled as a stellar wind with 
 a mass loss rate of the order of $6.6 \times 10^{-11}~ \msun ~\yr^{-1}$, which is 3,300 times the solar value,
 exceeding other estimates of this star's wind. However, interpretation of the emission in
 terms of other thermal mechanisms like coronal  free-free and gyroresonance emission cannot be discarded. 
\end{abstract}  


 
\section{Introduction}
At a distance of only 3.2 pc (van Leeuwen 2007) $\epsilon$ Eridani is one of the nearest stars to the
Sun. It has a spectral type of K2V (Keenan \& McNeil 1989) and an age that has been estimated to be
between 0.4 and 1.4 Gyr (Soderblom \& Dappen 1989; Soderblom et al. 1991; Janson et al 2008; Mamajek \& Hillenbrand 2008; Bonfanti et al. 2015).
The detection { of a debris disk} (Greaves et al. 1998; Holland et al. 1998; Chavez-Dagostino 2016)
located at $\sim$64 AU from the star with a width of $\sim$20 AU has stimulated many papers that study $\epsilon$ Eridani.

More recently, it has been found that there is continuum emission at millimeter and centimeter wavelengths associated with
the star (MacGregor et al. 2015; Chavez-Dagostino et al. 2016; Bastian et al. 2018). It has been suggested that the emission close to the star can {instead} be attributed to
one (or two) inner warm dust belts (Backman et al. 2009; Reidemeister et al. 2011),
although the results of Chavez Dagostino et al. (2016) indicate that the contribution
of such component might be marginal.

The centimeter continuum emission was detected and discussed in detail by Bastian et al. (2018). From radial velocity analyses of $\epsilon$ Eridani,
the presence of one giant exoplanet with semi-major axis of 3.4 AU ($\sim$1$''$) has been proposed (Hatzes et al. 2000, see also Mawet et al. 2018). The reality of this exoplanet, however,
remains controversial (Anglada-Escud\'e \& Butler 2012). In any case, Bastian et al. (2018) analyzed the quiescent and flaring radio continuum emission considering that
the K2V star or the Jupiter-like planet were the sources. While they could not rule out a planetary origin { based on the radio source position,} the nature of $\epsilon$ Eridani as a 
moderately active ``young Sun'' favors a stellar origin.

A direct way to distinguish between a stellar and a planetary origin is to obtain high angular resolution observations of the radio emission
and compare the radio position with the stellar position. A significant displacement between the two would favor a planetary origin, while a close coincidence will
favor a stellar origin. The previous radio observations of the emission associated with $\epsilon$ Eridani did not have sufficient
angular resolution to discriminate between the two possibilities. Here we present such observations. We also discuss the nature of the radio emission. {The broad bandwidth and high sensitivity of the available observations enable a detailed characterization of the spectral energy distribution, unprecedented for a young low-mass star.}

\section{Observations}

The observations were made with the Karl G. Jansky Very Large Array (VLA) of NRAO\footnote{The National 
Radio Astronomy Observatory is a facility of the National Science Foundation operated
under cooperative agreement by Associated Universities, Inc.} centered at a rest frequency of 33.0 GHz (9.1 mm) during
2017 June 15.  At that time the array was in its C configuration, providing an angular resolution of $\sim$1$''$.  
The phase center was at $\alpha(2000) = 03^h~ 32^m~ 54\rlap.^s84$;
$\delta(2000)$ = $-$09$^\circ~ 27'~ 29\rlap.{''}7$. The flux and bandpass calibrator was J0542$+$4951 and
the phase calibrator was J0339$-$0149. {The total observing time was 60 minutes, of which 28 were on-source. }

The digital correlator of the VLA was configured in 64 spectral windows of 128 MHz width divided 
in 64 channels of spectral resolution of 2 MHz. The total bandwidth for the 
continuum observations was about 8.0 GHz in a dual-polarization mode.

The data were analyzed in the standard manner using the CASA (Common Astronomy Software Applications) package of NRAO using
the pipeline provided for VLA\footnote{https://science.nrao.edu/facilities/vla/data-processing/pipeline} observations. 
Maps were made using a robust weighting (Briggs 1995) of 0 in order to optimize the compromise between sensitivity and angular resolution. 
The resulting image {\it r.m.s} was 9.0 $\mu$Jy beam$^{-1}$ at an angular 
resolution of $0\rlap.{''}85 \times 0\rlap.{''}51$ 
with PA = $-31.1^\circ$.
In Figure 1 we show the 33.0 GHz emission associated with $\epsilon$ Eridani.
The radio source appears unresolved with a flux density of 70$\pm$9 $\mu$Jy. The error given here includes only the statistical error. A more realistic
total error estimate would include a 10\% systematic error added in quadrature, giving a flux density of 70$\pm$11 $\mu$Jy.

Table \ref{table:1} summarizes the centimeter and millimeter observations of $\epsilon$ Eridani 
that will be discussed below. The first column is the instrument; the second column is the epoch of observation;
the third column is the frequency band; the fourth column is the flux density; the fifth column is an upper limit to the circular polarization; and the sixth column 
is the reference.

\begin{deluxetable}{l c c c c c}
\tablecolumns{4}
\tablewidth{0pc}
\tablecaption{Flux Densities of $\epsilon$ Eridani}
\tablehead{                        
\colhead{   }                    &
\colhead{   }                    &
Band &
Flux Density      &
$\rho_c^b$ &
  \\                
Instrument                           &
Epoch                   &
(GHz) &
($\mu$Jy)      &
($\%$)   &
Reference 
}
\startdata
VLA    & 2016-Mar-01 &   2--4 &  $<$65$^a$  & \nodata & Bastian et al.  2018 \\
VLA    & 2016-Jan-21 &   4--8 &   83$\pm$16.6  & $<$50 & Bastian et al.  2018 \\
VLA    & 2013-May-18 &   8--12 &   66.8$\pm$3.7  & $<$14 & Bastian et al.  2018 \\
VLA    & 2013-May-19 &   8--12 &   70.3$\pm$2.7  & $<$10 & Bastian et al.  2018 \\
VLA    & 2013-Apr-20 &   12--18 &   81.2$\pm$6.6  & $<$20 & Bastian et al.  2018 \\
VLA    & 2017-Jun-15 &   29--37 &   70$\pm$11  & $<$22 & This paper \\
ATCA    & 2014-Jun-25 to Aug-05 &   42-46 &   66.1$\pm$8.7  & \nodata & MacGregor et al. 2015 \\
SMA    & 2014-Jul-28 to Nov-19 &   217-233 &   1060$\pm$300  & \nodata & MacGregor et al. 2015 \\
IRAM   & 2007-Nov-16 to Dec-04 & 210-290   &   1200$\pm$300  & \nodata & Lestrade \& Thilliez 2015  \\
ALMA   & 2015-Jan-17 to Jan-18 & 226-234   &   820$\pm$68  & \nodata & Booth et al. 2017  \\
LMT   & 2014-Nov-01 to Dec-31 & 245-295   &   2300$\pm$300  & \nodata & Ch\'avez-Dagostino et al. 2016
\enddata
\label{table:1}
\tablecomments{
               (a): 2.5$\sigma$ upper limit for quasi-steady emission. \\
              (b): 2.5$\sigma$ upper limit for percentage of circular polarization.
            }
\end{deluxetable}

\subsection{Field Sources at 10 GHz}

Chavez-Dagostino (2016) detected a total of seven sources at 1.1 mm in a field of $\sim3' \times 2.5'$ centered on 
$\epsilon$ Eridani. To search for centimeter counterparts to these sources, we also reduced and analyzed the 3 cm (10 GHz)
Jansky VLA data taken as part of project 13A-471. These observations were taken on 2013 April 20, May 18 and May 19 with the VLA in the
D configuration. The data were calibrated and concatenated to
produce a deep image with a noise of $\sim$3 $\mu$Jy at the center of the field. We detected a total of 16 sources whose positions, flux densities and counterparts are
given in Table \ref{table:2}. Only three of these sources have previously known counterparts.
VLA source 2 coincides positionally with the galaxy 6dFGS gJ033243.6-093557 (Jones et al. 2009). 
VLA source 6 coincides with  $\epsilon$ Eridani. Finally, VLA source 11 coincides within $\sim$2$''$ with millimeter source 5 in the list
of Chavez-Dagostino et al. (2016).


\pagebreak

\begin{deluxetable}{l c c c c}
\tablecolumns{4}
\tablewidth{0pc}
\tablecaption{Sources Detected at 10 GHz}
\tablehead{                        
\colhead{   }                    &
\multicolumn{2}{c}{Position$^a$} &
\colhead{   }                    &
\colhead{}      \\
   &
$\alpha_{2000}$          &
$\delta_{2000}$          &   
Flux Density$^b$      &
  \\                
Number                            &
(h m s)                     &
($^\circ$ $^{\prime}$  $^{\prime\prime}$)       &   
($\mu$Jy) &
Counterpart 
}
\startdata
1$^c$    & 03 32 30.97 &   $-$09 24 30.9 &  3600$\pm$500  & \nodata \\
2$^c$    & 03 32 43.53 &   $-$09 35 56.7 &  240$\pm$40  & 6dFGS gJ033243.6-093557 \\ 
3    & 03 32 47.00 & $-$09 28 23.5 & 67$\pm$8 & \nodata \\
4    &  03 32 48.87 & $-$09 27 11.5 & 53$\pm$8 & \nodata \\
5   &  03 32 49.36 & $-$09 28 16.7 & 269$\pm$9 & \nodata \\
6   &  03 32 54.99  & $-$09 27 29.4 & 65$\pm$ 4 & $\epsilon$ Eridani \\
7   &  03 32 55.17  & $-$09 28 23.1 & 177$\pm$6 & \nodata \\
8   &  03 32 56.15  & $-$09 25 30.0 & 54$\pm$10  & \nodata \\
9 & 03 32 58.03 & $-$09 24 13.6 & 283$\pm$34  & \nodata \\
10   &  03 32 58.17  &  $-$09 25 06.1 & 1309$\pm$25 & \nodata \\
11 & 03 32 59.18 & $-$09 28 28.0 & 60$\pm$8 & S5 \\
12 & 03 32 59.90 & $-$09 27 50.8 & 124$\pm$23 & \nodata \\
13 & 03 32 59.94 & $-$09 27 52.6 & 42$\pm$7 & \nodata \\
14 & 03 33 08.77 & $-$09 28 09.8 & 291$\pm$34 & \nodata \\
15 & 03 33 09.21 & $-$09 28 43.6 & 500$\pm$45 & \nodata \\
16 & 03 33 10.66 & $-$09 28 57.5 & 647$\pm$90 & \nodata
\enddata
\label{table:2}
\tablecomments{
               (a): Positional errors are $\sim0\rlap.{''}2$.\\
              (b): Corrected for the primary beam response. \\
              (c): These two sources are far from the phase center and its flux density is a crude estimate.}
\end{deluxetable}

\pagebreak

\section{Comparison of radio and optical positions}

\begin{figure}
\centering
\vspace{-2.8cm}
\includegraphics[angle=0,scale=0.8]{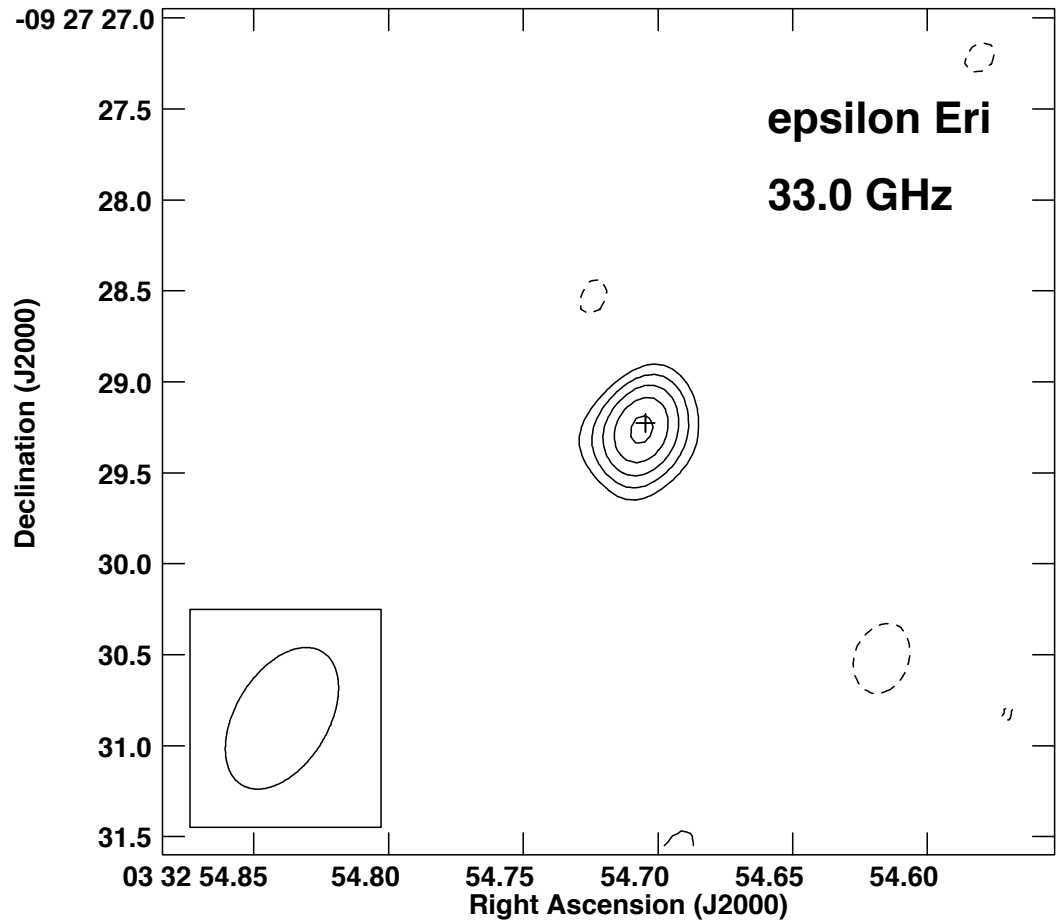}
\vskip-4.5cm
\caption{\small VLA 33.0 GHz continuum 
emission associated with $\epsilon$ Eridani.
The black contours are $-$4, $-$3, 3, 4, 5, 6, and 7 times 9 $\mu$Jy beam$^{-1}$, the rms of the image.
The half-power contour of the synthesized beam ($0\rlap.{''}85 \times 0\rlap.{''}51$ 
with PA = $-31.1^\circ$) is 
shown in the bottom left corner. 
The small cross indicates the position of the star, corrected for proper
motions and parallax (van Leeuwen 2007).
}
\label{fig1}
\end{figure}

We fitted the radio image with a Gaussian ellipsoid obtaining a position of
$\alpha(2000) = 03^h~ 32^m~ 54\rlap.^s7067 \pm 00\rlap.^s0022$;
$\delta(2000)$ = $-$09$^\circ~ 27'~ 29\rlap.{''}288 \pm 0\rlap.{''}032$.
The uncertainties in this position are the statistical noise resulting from the
least-squares fit.
To compare with this position, we used the Hipparcos position (van Leeuwen 2006) and corrected it
for proper motions and parallax to the epoch of the radio observations.
Taking into account the propagation of the astrometric errors, the Hipparcos data implies
$\alpha(2000) = 03^h~ 32^m~ 54\rlap.^s7046 \pm 00\rlap.^s0003$;
$\delta(2000)$ = $-$09$^\circ~ 27'~ 29\rlap.{''}228 \pm 0\rlap.{''}004$ at the epoch of the VLA observations.
This position is shown with a cross symbol in Figure \ref{fig1} overlaid on the radio emission. The difference between the Hipparcos and VLA positions are given by:

$$\Delta \alpha(\textrm{Hipparcos - VLA}) = -0\rlap.^s0021 \pm 0\rlap.^s0022$$

$$\Delta \delta(\textrm{Hipparcos - VLA}) = 0\rlap.{''}060 \pm 0\rlap.{''}032.$$

Since the separation between the Hipparcos and VLA positions are within 1 to 2 $\sigma$ of the positional error
we conclude that the positions can be considered to be coincident at the level of $0\rlap.{''}07$ (0.2 AU at the distance
of $\epsilon$ Eridani) and that this comparison rules out the possibility that the radio emission is coming from a position
displaced more than 0.2 AU from the star. The proposed giant exoplanet would have a semi-major axis of 3.4 AU (Hatzes et al. 2000) and thus would appear most of the time
to be displaced more than 0.2 AU from the star. We, therefore, support the suggestion of Bastian et al. (2018) that the
emission has a stellar origin. 
{ Note that, while we reach this conclusion on the basis of position coincidence, Bastian et al. (2018) reached the same conclusion from the fact 
that the observed radio emission is consistent with known mechanisms of stellar emission.}
\pagebreak

\section{The nature of the radio emission}

\begin{figure}
\centering
\vspace{-2.8cm}
\includegraphics[angle=0,scale=0.8]{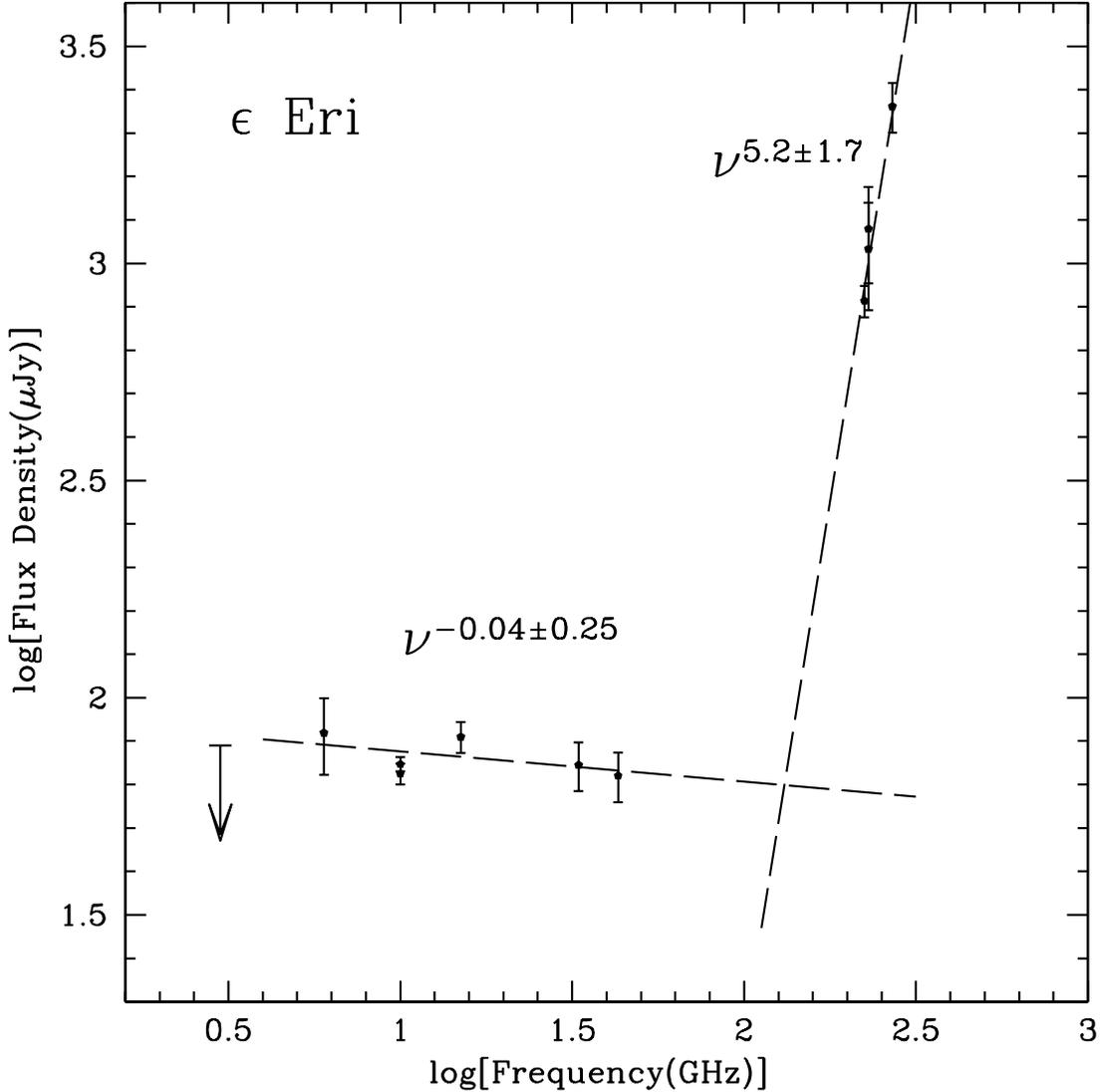}
\vskip-4.5cm
\caption{\small Spectrum of the emission associated with $\epsilon$ Eridani at cm and mm wavelengths from the references given in the text. The long-dashed lines are separate power law fits to
the cm and to the mm emission. The data point at 3 GHz (0.48 in the logarithm) is a 3-$\sigma$ upper limit and was not taken into account
in the fit. 
}
\label{fig2}
\end{figure}

In Figure \ref{fig2} we show the spectrum of the emission associated with $\epsilon$ Eridani in the centimeter and millimeter
ranges. These flux densities are taken from Bastian et al. (2018), MacGregor et al. (2015), Lestrade \& Thilliez (2015), Booth et al. (2017), Chavez-Dagostino
(2016) and this paper (see Table 2). 
In the case of the observed mm emission, the rapid rise of the flux density with frequency {(a spectral index of 4.9$\pm$1.8) favors emission from warm dust (e.g. Ricci et al.
2010).}
In contrast, the cm emission is remarkably flat { (a spectral index of -0.07$\pm$0.25)}, suggesting an optically-thin free-free nature. Below we examine the expected properties of the free-free emission.

\subsection{Optically thin free-free emission}

The free-free  power per unit volume per Hz is given by
\begin{equation}
e_\nu^{ff} = 6.8 \times 10^{-38} Z^2 n_e n_i T^{-1/2} \exp{(-h \nu/ k T)} \, \bar g_{ff}  \quad  { \rm erg \, s^{-1} \, cm^{-3} \, Hz^{-1} },
\end{equation}
where  $Z$ is the ion charge, $n_e$ and $n_i$ are the electron and ion number density, respectively, 
$T$ is the plasma temperature, and $\bar g_{ff}$ is the velocity averaged Gaunt factor.

The total power per unit volume is obtained integrating over frequency
\begin{eqnarray}
\epsilon_{ff} & = &  6.8 \times 10^{-38} Z^2 n_e n_i T^{-1/2}  \int_0^\infty \exp{(-h \nu/ k T)} \, \bar g_{ff}  d \nu, \\
&= & 1.415 \times 10^{-27} Z^2 n_e n_i T^{1/2} \bar g_B , 
\end{eqnarray}
where  $\bar g_B$ is the frequency average of the velocity average Gaunt factor, which is in the range 1.1-1.5.
Choosing $\bar g_B =1.2$ gives a 20\% accuracy (Rybicki \& Lightman 2004).

The power per unit volume over a frequency range is given by 
\begin{equation}
\epsilon^{\Delta \nu}_{ff} = 1.42 \times 10^{-27} Z^2 n_e n_i T^{1/2} \bar g_B \, (e^{-u_1} - e^{-u_2}),
\end{equation}
where the normalized energy is  $u= h \nu / kT$.
For an optically thin plasma at a constant temperature, the luminosity in a normalized energy range $(u_1, u_2)$  is obtained integrating the emissivity over the volume $V$
\begin{equation}
L_{\Delta \nu} = \int_V \epsilon^{\Delta \nu}_{ff} \, d V  = 1.42 \times 10^{-27} Z^2 T^{1/2} \bar g_B (e^{-u_1} - e^{-u_2}) \int n_e n_i d V .
\label{LDnu}
\end{equation}

The X ray luminosity of  $\epsilon$ Eridani has been measured with XMM-Newton in the energy 
range $[E_1, E_2] = [0.2, 2] $kev, with values in the range 1.5 $ -$ 1.68 $\times 10^{28}  \, {\rm erg \, s^{-1}}$ 
(Poppenhaeger et al. 2010; Lloyd et al. 2016). 
Let $L_R$ be the luminosity in the frequency range $[\nu_{1}, \nu_{2}] = [6, 43]$ GHz where the spectrum is flat.
Then, from eqn. (\ref{LDnu}), the 
 ratio of the free-free luminosity in the X ray
range  $L_X$ and in
radio range $L_R$  is given by
\begin{equation}
{L_X \over L_R} ={   e^{ -u_{1,X}} - e^{-u_{2,X} } \over  {  e^{ -u_{1,R}} - e^{-u_{2,R} }  }} = 3.5 \times 10^5  - 7.7  \times 10^5, 
\label{Lrat}
\end{equation}
where the lower value corresponds to  $T=2 \times 10^6$ K and the upper value corresponds to  $T=3 \times 10^6$ K.
These temperatures are  similar to those measured by Schmitt et al. (1996) for the peak of the temperature distribution of 
the coronal material of $\epsilon$ Eridani.
 %
 

The observed radio luminosity in the range 6 - 43 GHz is given by
\begin{equation}
L^{observed}_{R}  =  4 \pi D^2 \int_{6GHz}^{43 GHz} S_\nu d\nu  = 3.31 \times 10^{22} {\rm erg \, s^{-1}},
\end{equation}
where $D = 3.2$ pc is the distance to $\epsilon$ Eridani.
Then, this radio luminosity implies a free-free emission in X rays 
\begin{equation}
L_X = L^{obs}_{R}  \times {L_X \over L_R}  = 1.16  \times 10^{28}  - 2.57 \times 10^{28} {\rm erg s^{-1}}  , 
\end{equation}
for a temperature  in the range 2 $-$ 3  $\times 10^6$ K,
in agreement with the observed X ray luminosity observed by XMM-Newton. Therefore, the radio emission in the flat spectrum region
is consistent with free-free emission for this range of plasma temperature.
Note that line X-ray emission is also important in hot plasmas (e.g. Mewe et al. 1985; see their Figure 1).
Thus, the total X-ray emissivity should be larger than the continuum free-free emission calculated in this section.

There are at least two possible sources of the free-free emission: a magnetically-confined corona and a stellar wind. In the following section we will discuss the case of the stellar wind.

\subsection{Free-free Emission from a Stellar Wind}

In this section we explore the possibility that the radio emission comes from a stellar wind. We calculate the emission of an isothermal spherically 
symmetric fully ionized wind that has a spectrum  $S_\nu \propto \nu^{0.6}$ at low frequencies 
and  becomes optically-thin at high frequencies, such that $S_\nu \propto \nu^{-0.1}$ (e.g., 
Panagia \& Felli 1975; hereafter PF75).

Since the stellar wind is always opaque inside a radius that decreases with increasing frequency, 
one can estimate the turnover frequency of the stellar wind spectrum with the condition that the free-free optical depth is 1 
at an impact parameter equal to the radius of the star $R_*$.
The optical depth at a normalized impact parameter $\xi=p/R_*\ge  1$, where $R_*$ is the stellar 
radius, is given by eq. (4a) of PF75
$$
\tau(\xi) = {\pi \over 2 \xi^3}  n_*^2 R_* \kappa(\nu) ,
$$
where $n_*$ is the electron number density at $R_*$ and the opacity in the radio range  is given by
$
\kappa(\nu) = 8.436 \times 10^{-28} \left[{ \nu \over 10 GHz} \right]^{-2.1}   \left[{T_e \over 10^4 K}\right]^{-1.35} a(\nu, T_e),
$ 
where $T_e$ is the electron temperature and $a(\nu, T_e)$ measures the deviation from the exact 
formula (Mezger \& Henderson 1967).
The electron number density is given by  the mass continuity equation 
$
 n_*  = { \dot M_w  \over 4 \pi R_*^2 v_w \mu m_H },
$
where $\dot M_w$ and $v_w$ are the wind mass-loss rate and velocity, $\mu=1.2$ is the mean mass per electron, and $m_H$ is the proton mass.
Then, the turnover frequency of the stellar wind spectrum, given by the condition $\tau(1)=1$, is
\footnote{For simplicity, we assumed  $a(\nu, T_e) =1$  which has en error up to 20\% for
the range of parameters considered here.}
\begin{equation}
 {\nu_{to} \over 10~ \rm{GHz}}  = 1.19 \left[{R_* \over 0.74 \, R_\odot }\right]^{-1.42} \left[ {\dot M_w \over 10^{-11} \msun ~\yr^{-1}} \right]^{0.95} 
 \left[ {v_w \over 650\, \kms} \right]^{-0.95}  
 \left[{T_e \over 10^6 K}\right]^{-0.64} ,
\end{equation}
Furthermore, the level of the optically-thin emission can be estimated evaluating
the partially optically-thick flux in eq. (24) of PF75 at $\nu_{to}$ such that 
\begin{equation}
 {S_\nu^{\rm thin} } = 96 \mu Jy \left[{R_* \over 0.74 \, R_\odot }\right]^{-0.85}  \left[{\dot M_w \over 10^{-11} \msun ~\yr^{-1}}\right]^{1.9} 
 \left[ {v_w \over 650\, \kms} \right]^{-1.9}  
 \left[{T_e \over 10^6 K}\right]^{-0.29} \left[{d \over \, 3.2 ~{\rm pc}}\right]^{-2}  \left[{\nu \over \nu_{to}}\right]^{-0.1}.
 \end{equation}
This approximation has a percentage error $ < 25 \%$ for $\nu > 10$ GHz.

Although these equations are useful, to make a better comparison
with the observed radio emission, we obtain the wind spectrum by integrating numerically the transfer equation
$I_\nu (\xi) = I_\nu^{*} \exp(-\tau) + B_\nu(T_e) (1 - \exp(-\tau))$, where  the first term $I_\nu^{*} \exp(-\tau) $, 
included only for impact parameters $\xi \leq 1$, is the stellar specific intensity that is attenuated by the wind in front of it. 
The flux is obtained integrating $I_\nu(\xi)$ over the source solid angle, for frequencies in the partially optically-thick regime
to frequencies beyond the transition to the optically-thin regime.
\footnote{For the numerical integration we use the exact formula for $\alpha(\nu, T_e)$.}
We assume $v_w = 650 \, \kms $, which corresponds
to the escape speed of $\epsilon$ Eridani that has a mass $M_*=0.83 \msun$, a radius $R_*= 0.74 R_\odot$,
and a temperature $T_* = 5,100$ K  (e.g., Bonfanti et al. 2015). 
Fixing the wind speed leaves only two parameters to fit the spectrum: the wind mass-loss rate $\dot M_w$ and the temperature $T_e$. 
We consider 3 wind models with different temperatures that have the observed level of optically-thin emission 
Models A, B, and C  have  $T_e =  10^6 , 10^5, 10^4$ K, respectively. The mass-loss rate $M_w$ of each model is
determined by the requirement of fitting the observed optically-thin flux.
Table \ref{table:3} shows the temperatures, mass-loss rates, and turnover frequencies of these models. 
The turnover frequency increases with decreasing temperature (see eq. [1]). 

\begin{figure}
\centering
\vspace{-2.8cm}
\includegraphics[angle=0,scale=0.8]{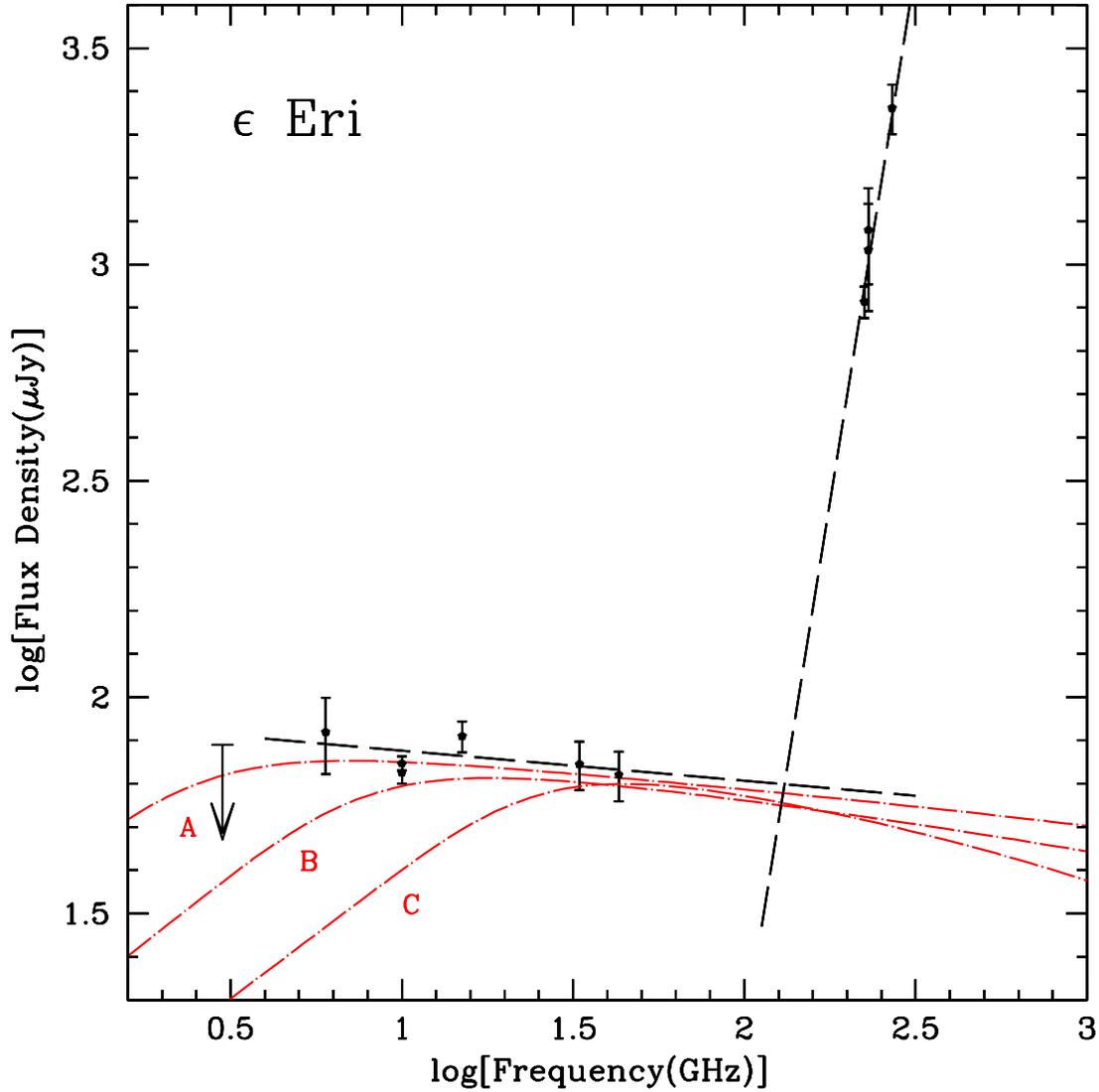}
\vskip-4.5cm
\caption{\small Spectrum of the emission associated with $\epsilon$ Eridani at cm and mm wavelengths shown in Figure \ref{fig2}. 
The red dashed-dotted lines show the ionized stellar wind models A, B, and C discussed in the text. 
Warm dust is required to reproduce the observed {\bf steep-spectrum} mm emission.}
\label{fig3}
\end{figure}

Figure \ref{fig3} shows the wind spectrum of Models A, B, and C with red dot-dashed lines superimposed on the data points. 
One can see that models B and C  do not fit the observed emission at 6 GHz because 
their turnover frequency is too high.  Therefore, one requires a wind with an electron temperature  $T_e \sim 10^6  K$.
The wind temperature of $\epsilon$ Eridani could be better constrained by determining the turnover frequency,  which should
be at frequencies  $ \nu < 6$ GHz.  

As discussed in the previous section, using eqn. (\ref{LDnu}) one
can calculate the X ray emission of this stellar wind and compare with the X ray luminosity
in the energy range $[0.2, 2]$ kev. 
 For a stellar wind where the density $n ={\dot M_w  \over 4 \pi \mu m_H v_w r^2} $, the X-ray luminosity in the energy range $[0.2, 2]$ kev is given by 
\begin{eqnarray}
{L_X \over {\rm erg \, s^{-1}}}
& = & 1.85 \times 10^{-27} T_e^{1/2} \left( e^{ -u_{1,X}} - e^{-u_{2,X} } \right) \int_{R_*}^\infty n^2 4 \pi r d r  \nonumber \\ 
& \sim  & 2.09  \times 10^{28}\,  e^{ -u_{1,X}} \left[{T_e \over 3 \times 10^6 K}\right]^{1/2} \left[ {\dot M_w \over 6.6 \times 10^{-11} \msun \yr^{-1}}\right]^2 \left[ {v_w \over 650~ \kms }\right]^{-2},
\end{eqnarray}
where $ e^{-u_{1,X}} =e^{ -0.77(E_1/ 0.2 ev) (T_e/3 \times 10^6)^{-1} }  = 0.46$.
A stellar wind with a temperature of $10^6$ K (model A in Figure 3) would produce half of the observed X-ray emission. \footnote{An electron temperature of $T_e \sim 10^6$ K, is in the range of the values 
measured by Schmitt et al. (1996) for the coronal material of $\epsilon$ Eridani.}
{\bf The other half may be due to line emission.}

As shown in Table \ref{table:3}, { assuming a spherical wind},
\footnote{The mass-loss rate can 
be overestimated by if the wind is collimated in a conical jet  because, for the same mass-loss rate, the density is higher in the latter case (e.g., see eq. (21) of Reynolds 1986). }
the  mass-loss rate required to reproduce the observed optically-thin radio emission of 
$\epsilon$ Eridani  is $\dot M_w \sim 6.6 \times 10^{-11} \msun \yr^{-1}$,   similar to the 
upper limits found by Fichtinger et al. (2017) for 4 young main-sequence solar-type stars using VLA and ALMA observations.
This mass-loss rate  is 3,300 times the solar value $\dot M_\odot = 2 \times 10^{-14} M_\odot {\rm yr}^{-1}$, 
and 110 times larger than the value obtained by Wood et al. (2005) for $\epsilon$ Eridani using the atmospheric Ly$\alpha$ absorption method. Nevertheless, the Ly$\alpha$ absorption method is indirect since it measures the heated HI within 
the interaction region between the stellar wind and the local ISM. In contrast, if the observed radio continuum emission is bremsstrahlung 
radiation, it is produced directly by  the ionized circumstellar gas. We note that models of winds from cool stars driven by Alfven waves and turbulence
by Cranmer \& Saar (2011)  predict a much lower mass-loss rates of the order of $\sim 5 \times 10^{-14} M_\odot {\rm yr}^{-1}$.
Other authors have  associated the energy of stellar flares  with Coronal Mass Ejections (CMEs)  with different assumptions to obtain the stellar mass-loss rate
(e.g.,Osten \& Wolk 2015; Odert et al. 2017). In particular, Odert et al. obtain a low mass-loss rate for $\epsilon$ Eridani consistent with the Ly$\alpha$ method. 
Nevertheless, Osten \& Wolk  obtain a mass-loss rate of the order of $4 \times 10^{-11} M_\odot {\rm yr}^{-1}$
for the young Solar Analog EK Dra that is similar to our value for $\epsilon$ Eridani.

In addition, the large mass-loss rate we find could help explain the Faint Young Sun Paradox,  in  which 
the geological evidence shows that the Earth and Mars had a much warmer climate in the past
(e.g., Feulner 2012).
The solution 
proposed by Whitmore et al. (1995) is that the Sun was more massive and luminous in the past but lost its excess mass 
in a solar wind with a mass-loss rate $\sim$1,000 times the present solar mass-loss rate, as large as the value we infer for $\epsilon$ Eridani. If the mass-loss rate remains constant,  $\epsilon$ Eridani would lose $\sim 0.2 ~\msun$ by the time it reaches the age of the Sun, although one expects
a decrease of the mass-loss rate with time (see, e.g., discussion of Fichtinger et al. 2017). 
 

\begin{deluxetable}{llll}
\tablecolumns{4}
\tablewidth{0pc}
\tablecaption{Ionized stellar wind models  }
\tablehead{
 \colhead {} & \colhead{$T_e$} & \colhead{$\dot M_{w} $} &  \colhead{$\nu_{to} $}\\
       {Model}                      &  (K)                      & $( M_\odot ~\yr^{-1})$         & (GHz) 
  }
\startdata
A & $ 10^{6}$ & $6.6 \times 10^{-11} $ &  3.71 \\ 
B & $10^{5}$  &  $4.2 \times 10^{-11} $ &  11.1  \\
C  & $10^{4}$ &  $3.0 \times 10^{-11} $ &  33.8  
\enddata 
\label{table:3}
\tablecomments{For all the models we assume $v_w = 650 \, \kms$}
\end{deluxetable}

\pagebreak

\subsubsection{Coexistence of the wind and disk}

{ A relevant question one could ask is:} can this relatively powerful wind blow out the dust grains in the surrounding debris disk?
To investigate this point we will make a rough comparison between the attractive force of gravity and
the repulsive force produced by the ram pressure of the wind, both acting on a dust grain.

The force of gravity will be given by
\begin{equation}
F_g = {{G M_* (4/3)\pi a^3\rho_d}\over{d^2}},
\end{equation}
where $G$ is the gravitational constant, 
$a$ is the radius of the dust grain, $\rho_d$ is the mass density of the dust grain
and $d$ is the distance from the grain to the star. On the other hand, the force due to the wind will be given by
\begin{equation}
F_w = {{\dot M_w v_w \pi a^2}\over {4 \pi d^2}}.
\end{equation}
The ratio between these two forces is given by
\begin{equation}
{{F_g} \over {F_w}} = {{16} \over {3}} {{GM_* \pi a \rho} \over {\dot M_w v_w}} \sim 180,
\end{equation}
where we adopted the stellar and wind parameters of $\epsilon$ Eridani discussed above, 
and assumed $a$ = 135 $\mu$m (Backman et al. 2009) and $\rho_d$ = 2 g cm$^{-3}$ (Love
et al. 1994),
Thus, the gravitational force dominates and the coexistence of a strong stellar wind with the debris disk is plausible. Other authors have studied the problem of the dust survival in debris disks  in more detail including radiation 
pressure, the Poynting-Robertson drag, and the stellar wind pressure (e.g., Plavchan et al. 2005; Augereau \& Beust 2006;  Strubbe \& Chiang 2006; Backmann et al. 2009). 
{\bf In the case of strong stellar winds this agent is more important for grain removal than radiation pressure and the Poynting-Robertson drag. }

\subsection{Other emission mechanisms }

Bastian et al. (2018) found that the radio emission in the range  $[4,12]$ GHz is quasi-steady since they did not find significant variability during the observations. 
They showed that the radio luminosity per unit frequency 
of $\epsilon$ Eridani in the 4 $- $ 8 GHz band is $l_R \sim 10^{12} \,  {\rm erg s^{-1} Hz^{-1}}$, therefore,   the ratio of 
$L_X/l_R \sim 10^{16.5}$ Hz. This is higher than the expected ratio  
of soft X rays and non-thermal radio spectral  luminosity
near 5 GHz for magnetically active stars,   $L_X/l_R \sim 10^{15.5}$ Hz, known as the 
G\"udel-Benz relation (e.g., G\"udel et al. 1995). Thus, $\epsilon$ Eridani is underluminous in the radio relative to active
stars. Bastian et al. (2018) argued that, even though 
 $\epsilon$ Eridani is younger and more active than the Sun, nonthermal radio emission is not required to explain
 the quasi-steady radio emission.
 In fact, as shown in Table \ref{table:1},  there is no evidence of 
circular polarization with upper limits of 12 $-$ 50\%. {\bf Detection of significant circular polarization (a few tens of percent) would favor a non-thermal gyrosynchrotron emission mechanism.}
Instead, Bastian et al. proposed that the  radio emission could be produced by thermal coronal 
free-free emission \footnote{Consistent with our interpretation of the emission as free-free discussed in Section 4.1.}
and thermal gyroresonance emission. A combination of these two mechanisms could explain the observed flat  radio 
 spectrum: the coronal free-free emission 
would dominate at the lower frequencies, while the gyroresonance emission would dominate at the higher frequencies. 
 We refer the reader to Bastian et al. (2018) for a detailed discussion of these mechanisms. 
{\bf We note that optically-thin free-free from $10^6$ K material (a stellar wind and/or magnetically confined corona) is a single mechanism that accounts for the flat spectrum in the whole observed range.}

\section{Conclusions}

Our main conclusions can be summarized as follows.

1) Using the VLA we detected 33.0 GHz emission associated with the nearby star $\epsilon$ Eridani. The radio and stellar positions coincide within
{\bf $0\rlap.{''}07 \pm 0\rlap.{''}05$ }and strongly favor the conclusion that the radio emission comes from the star and not from a proposed giant exoplanet with a semimajor
axis of $\sim$ 1$''$. 

2) The centimeter emission directly associated with $\epsilon$ Eridani is remarkably flat and  we show that it can be interpreted as
optically-thin free-free emission. We also show that this emission can be due to a stellar wind with a mass loss rate of order
$M_w \sim 6.6 \times 10^{-11}$ $M_\odot$ yr$^{-1}$. However, as discussed above, other mechanisms could also produce this emission, thus,
additional observations are needed to establish its nature. For example, the detection of circular polarization, the appearance of non-thermal
(that is, strongly negative) spectral indices or fast (in the scales of minutes) temporal variations will favor mechanisms other than optically thin free-free.

3) Sensitive 10.0 GHz observations made with the VLA reveal the presence of 15 sources in the field, in addition to the one associated
with $\epsilon$ Eridani. Only one of the 10.0 GHz sources appears to be associated with one of the 1.1 mm continuum sources
detected by Chavez-Dagostino et al. (2016).

\acknowledgements
We thank an anonymous referee for a careful revision of our paper that improved its clarity.
LFR, SL and LL acknowledge the financial support of
PAPIIT-UNAM IN105617, IN101418, IN112417, and CONACyT  23863. 
This research has made use of the SIMBAD database,
operated at CDS, Strasbourg, France.

\facility{VLA}

\software{CASA McMullin et al.(2007)}

\end{document}